# Full Wafer Redistribution and Wafer Embedding as Key Technologies for a Multi-Scale Neuromorphic Hardware Cluster

Kai Zoschke*[1], Maurice Güttler*[2], Lars Böttcher[1], Andreas Grübl[2], Dan Husmann[2], Johannes Schemmel[2], Karlheinz Meier[2], Oswin Ehrmann[3]

[1]Fraunhofer IZM, Gustav-Meyer-Allee 25, 13355 Berlin
[2]Heidelberg University, Kirchhoff-Institute for Physics, Im Neuenheimer Feld 227, 69120 Heidelberg
[3]Technical University of Berlin, Gustav-Meyer-Allee 25, 13355 Berlin

**Abstract**

Together with the Kirchhoff-Institute for Physics the Fraunhofer IZM has developed a full wafer redistribution and embedding technology as base for a large-scale neuromorphic hardware system. The paper will give an overview of the neuromorphic computing platform at the Kirchhoff-Institute for Physics and the associated hardware requirements which drove the described technological developments.

In the first phase of the project standard redistribution technologies from wafer level packaging were adapted to enable a high density reticle-to-reticle routing on 200 mm CMOS wafers. Neighboring reticles were interconnected across the scribe lines with an 8 µm pitch routing based on semi-additive copper metallization which was photo defined by full field mask aligning equipment. Passivation by photo sensitive benzocyclobutene (BCB) was used to enable a second intra-reticle routing layer. Final IO pads of nickel with flash gold were generated on top of each reticle. For final electrical connection the wafers were placed into mechanical fixtures and the IOs of all reticles were touched by elastomeric connectors. With that concept neuromorphic systems based on full wafers could be assembled and tested. The fabricated high density inter-reticle routing revealed a very high yield of larger than 99.9 %.

In order to allow an upscaling of the system size to a large number of wafers with feasible effort a full wafer embedding concept for printed circuit boards was developed and proven in the second phase of the project. The wafers were thinned to 250 µm and laminated with additional prepreg layers and copper foils into a core material. A 200 mm circular cut was done into the core material and the inner prepreg layers to create the required clearance for the wafer. After lamination of the PCB panel the reticle IOs of the embedded wafer were accessed by micro via drilling, copper electroplating, lithography and subtractive etching of the PCB wiring structure.

The created wiring with 50 µm line width enabled an access of the reticle IOs on the embedded wafer as well as a board level routing. The panels with the embedded wafers were subsequently stressed with up to 1000 thermal cycles between 0 °C and 100 °C and have shown no severe failure formation over the cycle time.

## 1. Introduction

The EU funded Human Brain Project (HBP) [1] is developing methods and tools to provide neuroscience with state-of-the-art and future computing capabilities. The HBP is operating six research platforms of which one is dedicated to neuromorphic computing (NMC), a technology implementing brain-like architectures on electronic substrates. As part of the HBP NMC platform the BrainScaleS system is based on physical models of brain cells (neurons), their connections (synapses) and the underlying network structure. Physical models closely resemble biology as they are implemented with local analog circuitry for neurons and synapses and binary, event (spike) based communication between neurons. BrainScaleS is an accelerated physical model in which relevant electronic time constants are scaled down by a factor of 1000-10000 compared to biology with the goal to emulate various neural mechanisms of learning and development in a single experimental set-up. The acceleration causes extreme requirements on the communication bandwidth in the network, which have been met by employing wafer scale integration for the inter-chip connectivity.

The BrainScaleS base chips are mixed-signal ASICs. The first generation HICANN (High-Input-Count-Analog-Neural-Network) chip was manufactured in a 180 nm CMOS technology by the company UMC. It features 114,688 programmable dynamic synapses and up to 512 neurons [2]. A single 200 mm wafer carries 384 usable HICANN chips. The wafers are mounted into modules, which deliver electrical power and carry FPGA modules for the communication with a host computer. The FPGAs are used to configure the chips as well as to transmit events to and from the neural circuits on the wafer [3]. The HBP NMC platform currently comprises a total of 20 wafer modules from the first generation HICANN chips with approximately 1 Billion synapses and a maximum of 4 Million neurons. The second generation HICANN chips are currently realized in a 65 nm CMOS technology by the company TSMC. They offer several advanced computational features like an on-chip general purpose RISC processor for local learning [4] and a multi-compartment neuron model [5].

The full wafer redistribution and printed circuit board embedding described in this paper was carried out using the first generation wafers in 180 nm technology. The applied redistribution layer (RDL) technology was adapted from state-of-the-art processes for fan-in wafer level chip size packaging which are broadly used today [6, 7]. The full wafer embedding technology was adapted from a standard technology for chip embedding into printed circuit boards [8].

All process and technology developments were done with the goal to transfer the results also to the wafers with 65 nm technology for the second generation system, later. To enable a further upscaling, the processes are also suitable for a transfer to 300 mm wafers.

## 2. Current Hardware Platform based on Full Wafers with High Density RDL

The BrainScaleS system consists of 68 electronic sub-assemblies and 14 mechanical components. In Figure 1 (a) an exploded view of the system is shown. In the center is the HICANN wafer (A) with the mainPCB (D). Between both is a positioning mask with 384 elastomeric connectors (C). At the bottom are the 48 FPGA modules (B) and on top are the boards for the power supply of the system (E, F). When completely assembled the module occupies a space of 50x50x15 cm³. The module is designed for a worst-case power consumption of 2 kW.

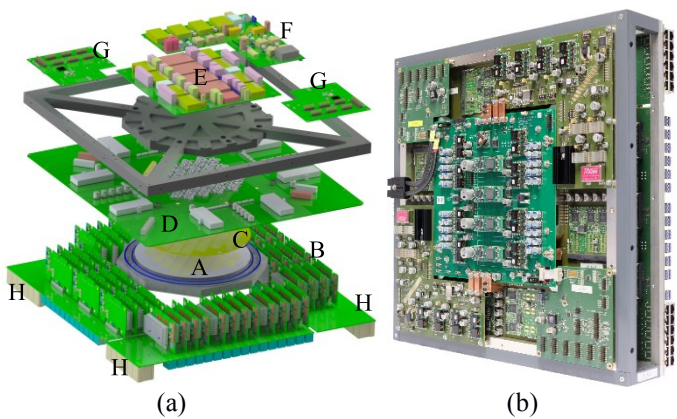

**Figure 1: (a) exploded view of the BrainScaleS system. See main text for explanation. (b) Photograph of a fully assembled wafer module.**

The mainPCB has a length and width of 43 cm. 14 copper layers distribute twelve supply voltages to the wafer, 1150 measurement lines and over 3000 data signals [9]. To facilitate signal routing, micro vias have been stacked over five copper layers to avoid penetrating the routing area on the other layers with through-hole vias. The 48 FPGA modules are used for the communication between the HICANN chips on the wafer and the host computer. For the host communication every FPGA module is equipped with Gigabit-Ethernet. 12 of these connections are routed to each edge of the module, respectively (H). On the wafer side one FPGA controls eight HICANN chips that together account for one reticle. Every HICANN has two full-duplex serial LVDS links with separate clock and data lines to the FPGA module. The link is capable of transmitting two GBit/s. In total the FPGAs handle 384 HICANN chips. The neuron model on the HICANN chip is implemented with analog electronic components and, therefore, runs in a continuous time mode. In contrast, the communication of neuron events uses digital packets [10]. As a result, the circuitry is power efficient and the complete HICANN consumes only 1.3 W/cm². Due to the accelerated operation of the analog circuitry delays in the communication lines are crucial to the neuronal network behavior. A mean firing rate of 10 Hz in the biological time domain transforms to 100.000 Hz in the electronic time. Moreover, the input count of synapses to a neuron can be increased by combining adjacent neurons. A maximum of 64 combined neurons leads to 14336 synapses per neuron. This means that on average 1.4 GEvents/s have to be transmitted per neuron. However, in periods of bursting neuron activity it can be much higher. The aforementioned constraints require a communication link with low latency and high through-put. Furthermore, it should have a low power consumption. This is achievable by keeping the silicon wafer as a whole and create connections between the chips on the wafer, in addition to LVDS interface. The advantage is shorter lines with a very high density and a lower capacitive load. A drawback is that the standard CMOS manufacturing process restricts the layout to the lithography area inside a reticle. In case of the wafers made in 180 nm technology this area is around 20 x 20 mm and eight HICANNs fit inside. Hence, a post processing of a wafer-scale RDL is required to create the connections across reticle borders. In total the on-wafer network connectivity requires 2048 lines across the horizontal edge and 1280 lines across the vertical edge for the interconnection of adjacent reticles. Additionally, a second RDL is required to add pads for the connection of the wafer to the mainPCB. Two HICANNs have a combined pad layout with 62 pads which sums up to 11904 pads on the complete wafer. The pad sizes vary depending on the expected current flow. The smallest pads (i.e. for the LVDS lines) are 1200x200 $\mu m^2$. Between the main PCB and the HICANN wafer 384 elastomeric connectors from Fujipoly establish the vertical connections. In Figure 2 the working principle of an elastomeric connector is depicted. They have conductive and insulating layers with a width of 50 μm. After compression to around 80 % of its initial height the silver balls in the conductive layer create an electrical connection from top to bottom.

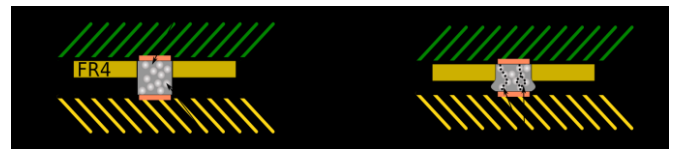

**Figure 2: Working principle of an elastomeric connector. The FR4 positioning mask keeps the elastomeric connector in its position.**

With the BrainScaleS system concept large neuronal networks on one HICANN wafer with 200.000 neurons and 40.000.000 synapses are possible. For network sizes larger than one wafer multiple modules can be interconnected. At the moment the neuronal events are routed over a compute cluster. Later, it is intended to connect the modules directly with each other.

## 3. Full Wafer High Density RDL Technology

As discussed in chapter 2 the usage of full wafers with interconnections between the reticles is the key approach for the implementation of the neuronal networks in this work. However, due to stepper based lithography in front end fabs an interconnection wiring across reticle borders could not be implemented in the back end of line (BEOL) stack of the HICANN wafers. For the establishment of an inter-reticle as well as additional intra-reticle wiring a post processing by wafer level packaging technologies is most appropriate. These technologies like wafer bumping or redistribution typically use full field lithography masks so that all areas on the entire wafer are exposed simultaneously during one single exposure step. Based on this, a structure definition between dies or reticles becomes possible.

For the realization of the required ultra-high density routing between the reticles of the HICANN wafers as well as for the regular intra-reticle routing the standard redistribution process at Fraunhofer IZM for fan-in chip scale packaging was adapted and optimized to meet the extended requirements. The standard redistribution process which is shown schematically in Figure 3 is based on deposition and structuring of four functional layers. Layer 1 is a thin photo sensitive polymer of 5 μm thickness which covers the entire wafer. Vias are structured into this polymer above each chip IO pad to be connected. The second layer is the redistribution wiring of 3-5 μm thick electroplated copper with line pitches not less than 40 μm. This wiring layer connects the original chip pads and re-routes them typically to an area array configuration with larger pad size and pitch. The third layer is a further polymer of 5-10 μm thickness which covers the wiring structure. It is only opened on top of the new created pad structures. The fourth layer is typically electroplated nickel with gold flash which is deposited as under bump metallization at the via openings of the second polymer layer.

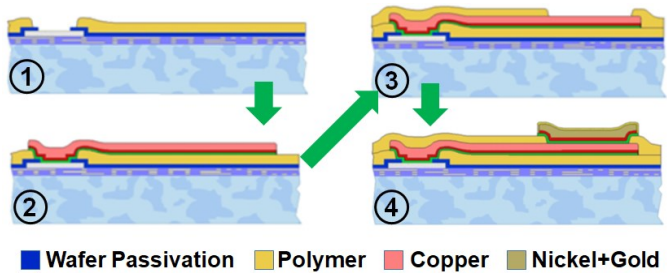

**Figure 3: Standard redistribution technology for fan-in chip scale packaging**

Based on earlier experiments with electroplated fine pitch routing over entire wafers a minimum possible line pitch of 8 μm was defined for the inter-reticle routing on the HICANN wafers. Considering estimations on possible full field mask alignment accuracy, required overlaps for capture pad to IO pad alignment as well as minimum possible via opening dimensions in the first polymer layer an optimized arrangement of the high density IO pads along the reticle borders could be defined.

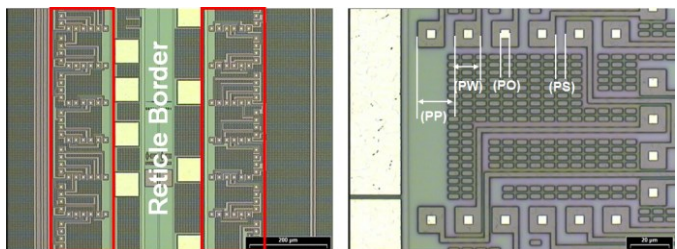

**Figure 4: Example image of the reticle border with the pad configuration for interconnection with ultra-high density RDL (left), image with closer zoom into pad configuration (right)**

The left image in Figure 4 shows an example of the IO pad arrangement along the reticle borders of a HICANN wafer as received from the wafer fab. The areas with the IOs are marked in red. The right image in Figure 4 shows a closer zoom to the high density IO pads. The pad pitch (PP) is 19 μm and the pad width (PW) is 15 μm. The pads are separated by a spacing (PS) of only 4 μm and the passivation opening (PO) on top of each pad is 5 μm.

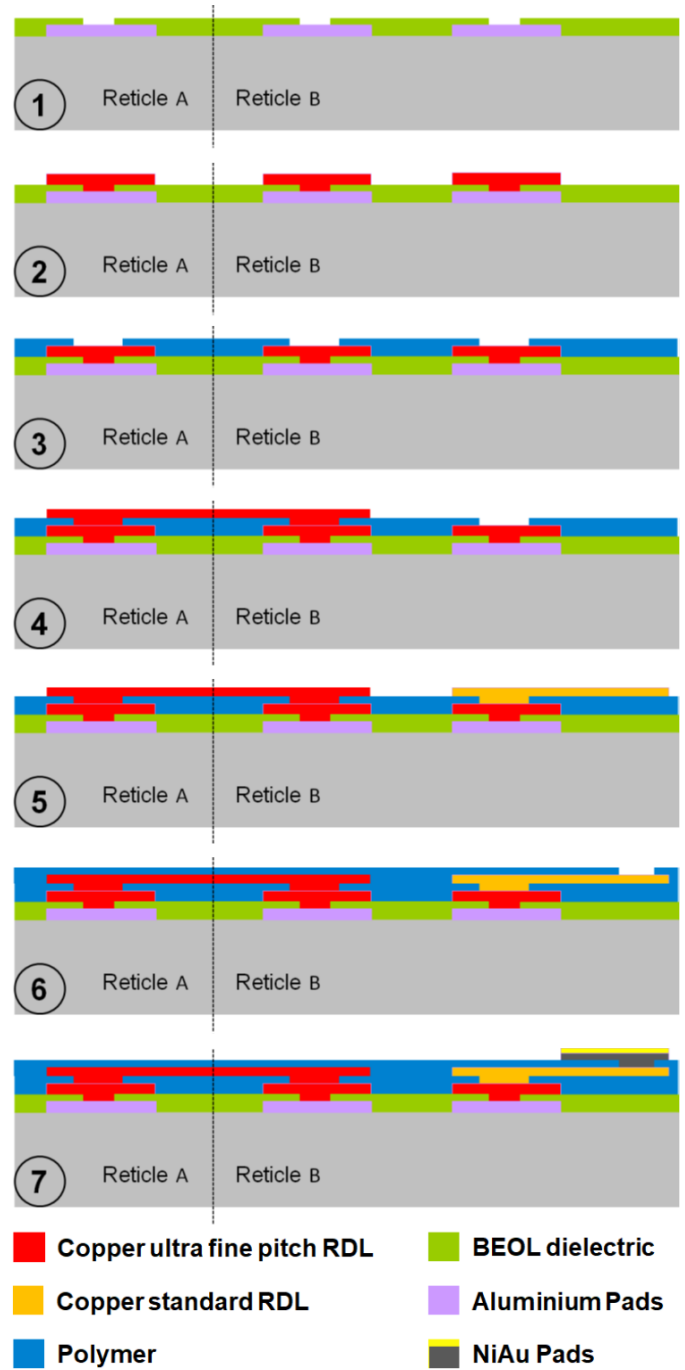

**Figure 5: Schematic process flow of adapted technology for full wafer redistribution with ultra-fine pitch inter-reticle routing**

The wafers with the described high density pad configuration were treated with the new developed post processing sequence which is schematically shown in Figure 5.

All lithography steps were done by using mask aligner equipment and full field 9 inch photo masks.

Picture 1 in Figure 5 indicates the initial state of the wafers as shown in Figure 4. Picture 2 in Figure 5 shows the result after finishing the first process sequence were copper posts with 4 µm height were created on top of each IO pad. The copper was deposited by semi-additive technology, which means, that the copper was electroplated into openings of a resist mask. The resist mask itself was structured on a sputtered adhesion and seed layer. The adhesion layer is also thick enough to act as diffusion barrier between the copper posts and the aluminum IO pads. After the electroplating step, the resist was removed and the seed and adhesion layers were etched away.

In the next process sequence (picture 3 in Figure 5) the polymer passivation layer was fabricated. Due to the small via geometries which had to be realized the usage of photo sensitive polymer and direct structuring by lithography as in the standard process was not possible here. Instead of this, the via openings were realized by dry etching using a photo resist as mask covering the polymer. Non-photo sensitive BCB was used here as a polymer. The dry etching process stopped on the surface of the copper posts after the polymer was etched through without damaging the original reticle IO pads. The copper posts allow an optical inspection of the etching result and thus a secure determination if the polymer was completely etched away.

Picture 4 in Figure 5 shows the result after the next process sequence were the actual ultra-fine pitch wiring across the reticle borders with 8 µm pitch was generated. The wiring is fabricated by semi-additive copper deposition with a height of 3 µm. The thicknesses of the sputtered adhesion and seed layer were reduced here to enable later their easy removal by wet chemical etching without the risk to remove too much material from the fine pitch RDL or to damage it by under etching. However, after removal of the photo resist the seed layer was not yet etched. Instead of this, an additional photo resist lithography was done to define a further wiring structure which runs inside the reticles and does not cross reticle borders. The so-called intermediate routing has relaxed line widths and spaces comparable with standard RDL features. The intermediate routing is indicated by the orange structure appearing in picture 5 of Figure 5. After deposition of the intermediate routing the seed and adhesion layers were wet chemically etched.

In the next process sequence a photo sensitive BCB polymer layer was deposited and structured by direct exposure and development. Due to the larger vias located in this layer the standard process from regular RDL technology could be used here. Picture 6 in Figure 5 shows the schematic cross section of the build up after the polymer process was finished.

In the final process sequence the nickel pads with gold flash were fabricated using semi-additive technology. The final schematic build-up of the wafers after completed post processing is shown in picture 7 of Figure 5. The created pads serve as interface to the next system level. An elastomeric connector as shown in Figure 2 will touch down on each individual pad to establish the electrical connection in the final system assembly.

The left image in Figure 6 shows the fabricated ultra-high density RDL structure after the complete post processing sequence was finished. The RDL routing runs across the reticle borders and provides interconnection between adjacent reticles. The image shows a similar area of the wafer as the left image in Figure 4 which shows neighboring reticles before the post processing was executed. The right image in Figure 6 shows a closer zoom into the connection area of the routing were the RDL is connected to the reticle IO pads. The image shows the same area on the wafer as the right image in Figure 4 which was taken before the post processing was performed. As indicated in Figure 6, the fabricated ultra-fine pitch routing has a line to line pitch (LLP) of 8 µm with a line width (LW) of 4 µm and a line to line spacing (LLS) of 4 µm. The spacing between the capture pads and the lines (PLS) is also 4 µm.

Figure 7 shows an image of one full HICANN wafer after completed post processing. The detail views show examples of the ultra-fine pitch RDL, as well as the intermediate RDL and the final nickel/gold pads for the touchdown of the elastomeric connectors. By using the described technology in total about 160.000 functional inter-reticle connections were established on each HICANN wafer. Electrical tests of the connections revealed a routing yield of >99.98 %.

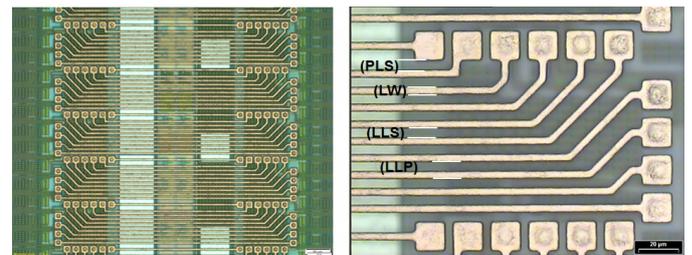

**Figure 6: Fabricated ultra-high density RDL across reticle borders (left), closer zoom into connection area of RDL to high density IO pads (right)**

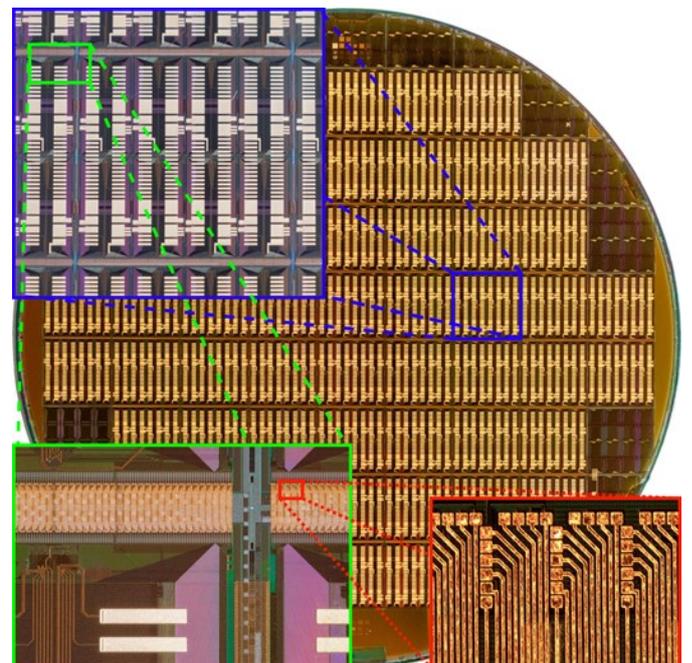

**Figure 7: HICANN wafer with RDLs. Enlargement views show a reticle and the fine-pitch lines across the reticle borders.**

## 4. Up Scaled Neuromorphic Hardware Cluster with higher Wafer Numbers using Wafer Embedding Technology

At the moment, 20 BrainScaleS systems are assembled and in operation in Heidelberg. They occupy five standard 19" server racks. So the distance between the neuromorphic modules is in the range of meters which adds a significant delay in the communication chain. Moreover, the complete assembly and testing of one system takes at least one workday. Most of the assembly steps are done by hand and are not automatable, e.g. filling of the positioning mask with elastomeric connectors and the alignment of HICANN wafer and mainPCB. Therefore, building a large-scale neuromorphic hardware cluster with the current BrainScaleS integration concept is limited.

For a large-scale neuromorphic cluster a new concept is required. A change to smaller process technologies does not necessarily increase the number of neurons and synapses. The next generation of HICANN chips in 65 nm technology will have approximately the same amount of neurons and synapses, with more sophisticated features. Additionally, in 65 nm technology more digital circuits can be placed in the same chip area which allows the integration of a general purpose RISC processor to implement local learning rules [4]. In the long run the external components of the BrainScaleS system could be implemented directly on the wafer which would reduce the volume of the system. For hundreds of systems the current wafer connection concept also has to be replaced. Therefore, it is intended to embed the silicon wafers into printed-circuit boards. With a reduced wafer thickness of 250 μm the total board thickness can be reduced to millimeters. The contacts through the laminate to the embedded silicon wafer are created by micro vias filled with copper, the production of the board uses standard PCB manufacturing techniques. As a consequence the amount of manual work is drastically reduced. The absence of external components allows the stacking of multiple boards which is schematically shown on the left side in Figure 8. The image shows a possible stack of five boards with water cooling modules in between to handle the heat dissipation.

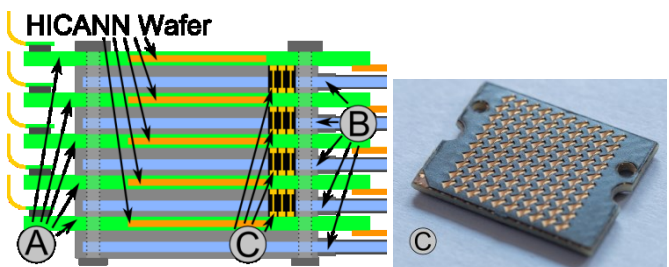

**Figure 8: Stacking of boards with embedded wafer systems (A) and water cooling module (B) (left), Samtec ZA1 connector (C) with dual compression contacts for multiple stacking of boards (right)**

The flat boards without mounted components allow a vertical stacking with direct interconnections between them. Instead of elastomeric connectors for the vertical connections interposers with compression contacts on both sides will be used. The picture on the right side in Figure 8 shows a Samtec ZA1 connector which has an array of 10x10 pins with a pitch of 1 mm. Due to the coarse pitch of the pins only a simple alignment of the boards is required.

## 5. Full Wafer Embedding Technology

Embedding of active and passive components into printed circuit boards (PCBs) meanwhile has moved from academicals and research activities to an industrialized process [8]. The technology is based on laminate (prepreg) embedding and PCB processing. The basic process flow is the component placement, e.g. to a FR4 core, embedding by vacuum lamination using epoxy prepreg, micro via laser drilling to the embedded components and a subsequent metallization and structuring of the wiring lines.

The idea of embedding a complete wafer into a PCB, in order to create the wiring substrate around, is based on this technology, although the "chip" becomes much larger. That's why warpage control of the substrates became of paramount importance. As illustrated in Figure 9, for embedding of the wafer, a suitable core is needed which is surrounding the wafer and acting as the substrate material. Initial trials did show that by using regular PCB materials only, warpage of the substrate could not be controlled. As a result, a handling and processing of the substrates was not possible.

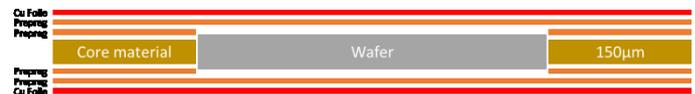

**Figure 9: Schematic build up wafer embedding with symmetric layup**

Due to that reason, different core materials were tested. Although ultra-low CTE (2 – 4ppm/K) laminate materials provide improved warpage control, the best results were achieved using a copper-invar-copper substrate (CIC). Invar is a nickel/iron compound with a CTE of around 1.2ppm/K, controlling the lateral extension of the substrates, and with that proving a good warpage control.

Based on the use of CIC as core substrate, an adapted process flow was developed, which is shown in Figure 10. To enable this flow the thickness of the HICANN wafers was reduced to 250 μm by back grinding and subsequent dry polishing.

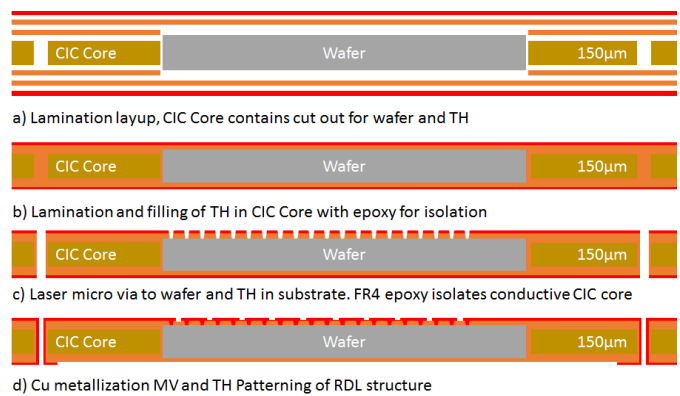

**Figure 10: Process flow for full wafer embedding**

In order to place the wafer in the core, first a cut out is made into the core material. For a double sided routing on the substrate, additionally through holes are drilled. Since the CIC compound is electrically conductive, they need to be isolated again later. For embedding, the lamination layup is applied. A fully symmetric build up is realized by applying identical FR4 prepreg layers on both sides of the core. The final layer is a thin copper foil, which is needed as base for the copper metallization. Image (a) in Figure 10 shows the schematic cross section of the lamination layup. Figure 11 illustrates the layup of the different layers. After preparing the layup the stack is transferred to a vacuum lamination press, were pressure and a temperature ramp is applied. During this process the resin of the prepreg material firstly starts to flow and fills all gaps and through holes (THs). In the second phase of the lamination, the epoxy resin is fully cured. Now the wafer is embedded into the substrate and the through holes are isolated with epoxy resin which is shown in picture (b) of Figure 10.

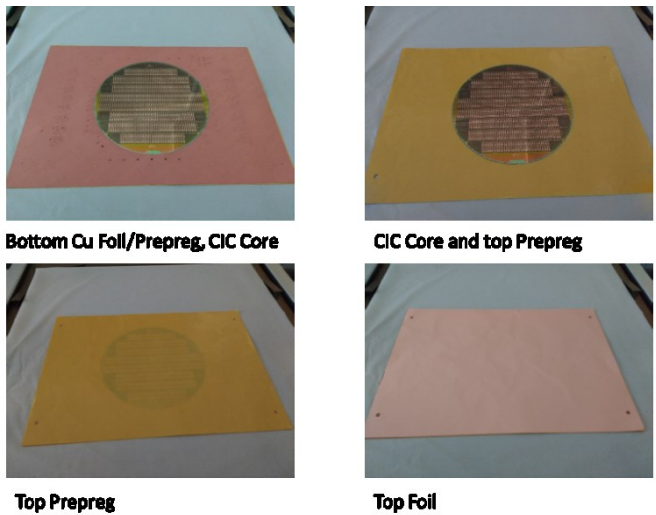

Figure 11: Layup of CIC core, epoxy prepreg layer and Cu foil

To realize the access to the contact pads of the embedded wafer, micro vias are made by UV laser drilling. The position of the vias are defined by Gerber data input. To enable the UV drilling with secure stop on the wafer contact pads they were reinforced by 12 µm thick electroplated copper during the wafer processing. The micro vias have a nominal diameter of 50 µm. According to the thickness of the prepreg layer the via depth is approximately 50 µm. Additionally to the micro vias, the prepared through holes in the CIC material are re-opened by laser drilling with a smaller diameter as the original holes. The remaining epoxy material in the hole is used as an isolation layer to the conductive CIC layer. This way vertical interconnects between front and backside of the substrate are realized which is shown in picture (c) of Figure 10. Finally, copper is deposited and the routing structure is patterned by lithography and etching as indicated in picture (d) of Figure 10. The routing layer has a thickness of 18 µm and a minimum line pitch of 80 µm. The minimum line width of the routing is 50 µm.

Figure 12 shows the resulting "Deep EvolutioN in System Embedding" (DENSE) board which was created. Using the described process it was possible to build an almost flat substrate with the embedded 200 mm HICANN wafer inside. The board has a total size of 360 x 240 mm$^2$. The thickness of the board is 0.42 mm.

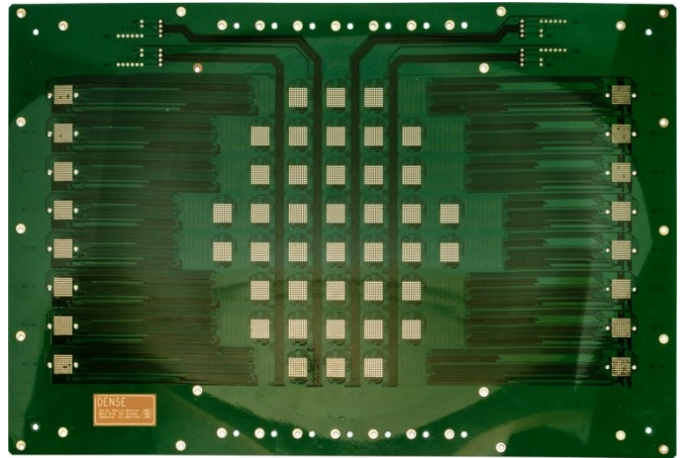

Figure 12: Top view on DENSE board with an embedded HICANN wafer. Each 10x10 pad array connects to two HICANN chips on the wafer.

## 6. Mechanical, Electrical and Reliability Characterization of Fabricated Prototypes

The test setup for the DENSE board uses an existing test setup which was developed for HICANN test chips. In principle, it is a small scale version of the BrainScaleS system. It uses the same FPGA boards and software stack as the wafer-scale system. In Figure 13 the prototype setup is shown. From a user perspective there is no difference in the use of the systems. In total up to eight HICANN chips can be connected to this setup.

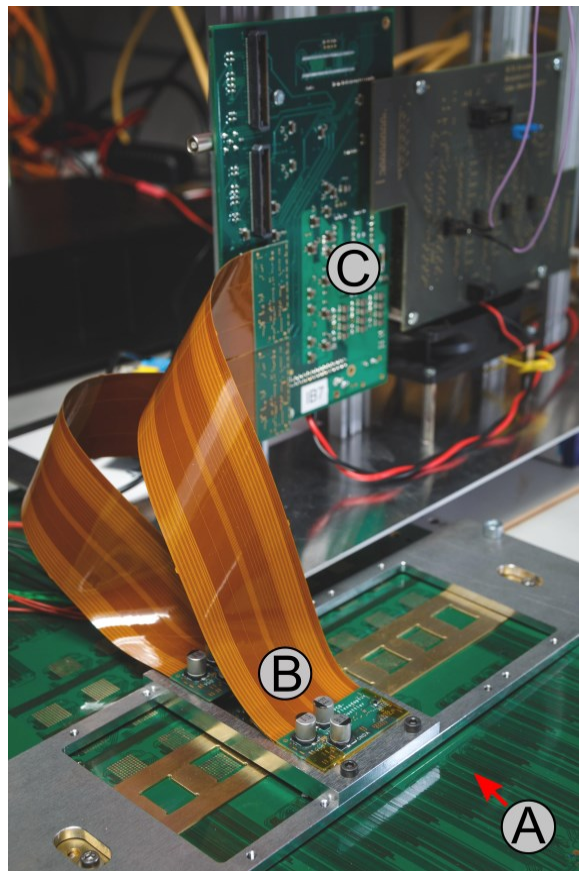

Figure 13: DENSE prototype setup. At the bottom is the DENSE board (A) which is connected with flex-rigid adapter boards (B) to the HICANN test setup (C).

The DENSE board has one copper layer per side, hence, only two HICANNs per reticle could be routed to a ZA1 connector. During the lamination process the wafer rotated slightly. However, the copper layout and the micro vias could be adjusted to the new position of the wafer pads. Unfortunately, the new positions of the plated through holes (PTHs) connecting top and bottom of the board were not in line with the pre-drilled holes in the CIC material. If the PTHs were added, there would be shorts. Hence, the PTHs were not created that reduced the number of accessible HICANNs from 112 to 80 chips. These are equally distributed over the wafer. For preventing the rotation issue to occur again wafers for next embedding runs will be equipped with two flats along their edges.

The characterization of the embedded wafer was done with three tests. First the communication over a JTAG port and LVDS links is tested. This is done by writing random data in Static Random Access Memory on the chip and comparing the sent value with the read back value. The second test checks the proper functioning of analog parameter storage circuits on the wafer. The parameters are stored in floating-gate cells. The storage consists of 4 blocks with 24 rows and 129 columns each which adds up to 12384 floating-gate cells per HICANN. Every cell can be programmed individually to a voltage between 0 V and 1.8 V. Every cell is programmed with a random value and then checked by reading the analog voltage. In the last test a neuron calibration is conducted on all 512 neurons. The calibration software tries to find neuron parameters such that the neurons behave like the corresponding numerical neuron model and eliminates fixed pattern noise between the neuron circuits. The number of rejected neurons is used for the characterization of the embedded wafer system. For these neurons no functioning parameter set could be found.

In summary, it can be said that none of the described tests showed a problem which could be related to the embedding process [11]. On the DENSE board the communication worked flawlessly with 77 of the 80 HICANN chips. One HICANN chip had completely broken floating-gate blocks which could not be programmed to the desired values. The values were either 0 V or 1.8 V which indicates to a problem in the floating-gate controller circuit. Therefore, the neurons on this chip could not be calibrated too. Apart from that there was no further broken floating-gate block detected. The neuron calibration revealed another HICANN chip where almost all neurons could not be calibrated. The reason was most likely a broken circuit block which affected all neurons on the chip. These errors also occur in the BrainScaleS system and for the remaining chips the number of sorted out neurons was in the expected range. In Figure 13 is the result for neuron calibration depicted. There are no influences of the embedding process visible on the chips and the RDLs.

In order to approve the robustness of the embedding technology as well as the reliability of the micro via connections and board level RDL accelerated environmental stress tests were performed with embedded dummy wafers. Those wafers had the identical multi-layer post processing as the functional wafers including IO pads at the top level for landing of the micro vias. A special routing layout was created for the embedding of these dummy wafers to establish daisy-chains which run alternating at the wafer and at the board. Those test PCBs were exposed to temperature cycling between 0 °C and 100 °C using a cycle time of 30 min. Up to 1000 cycles were executed. The resistance values of 15 daisy chains were observed during interruptions of the cycling load as well as after all 1000 cycles. All 15 chains were still fully intact after completion of the cycling tests.

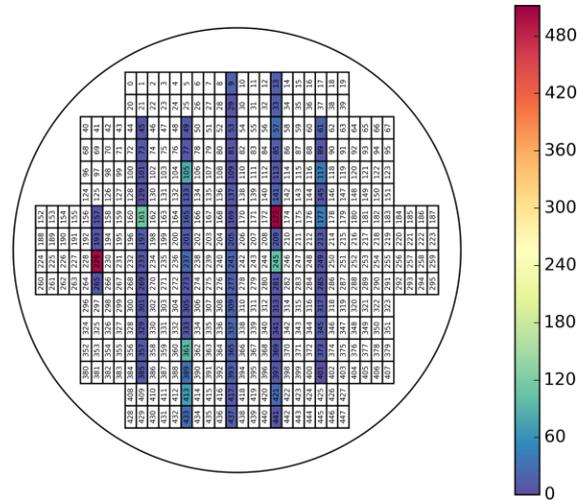

Figure 13: Wafer map with the number of sorted out neurons by the calibration software for the DENSE board.

A DENSE board with an embedded functional wafer had to undergo an accelerated environmental stress test as well. It was placed in a climate cabinet and put under thermal stress. The temperature changed cyclic between 15 °C and 90 °C. The real operating temperatures of a system are between 25 °C and 60 °C. In total 393 thermal cycles were driven. After that the board was placed in a soldering furnace to see a possible influence of the soldering process. The furnace reaches a maximum temperature of 230 °C. An optical inspection showed no delamination of the stack up or warpage of the board. A subset of the available chips on the embedded wafer was checked with the same tests as before. No failures induced by the thermal stress could be observed. All connections to the wafer worked fine.

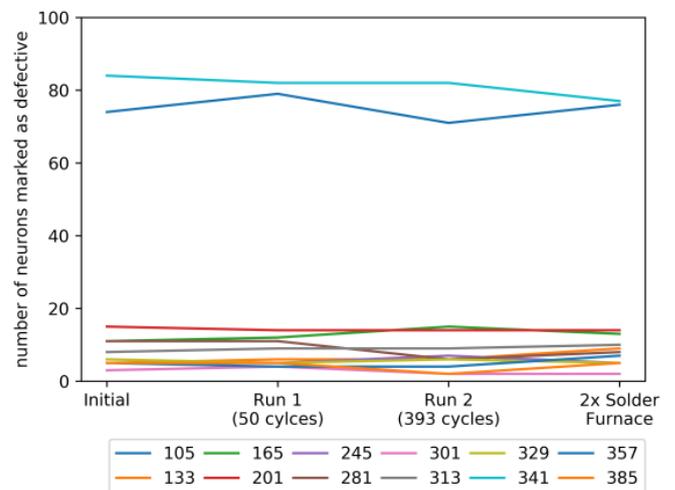

Figure 14: Temporal changes of the results of the neuron calibration during the thermal stress tests.

In Figure 14 the temporal change of the number of sorted out neurons is plotted. The changes between the test runs are in the range of trial-to-trial variations.

In the prototype setup the Samtec ZA1 connectors are used as a solderless connection to the DENSE board. The ZA1 has to be compressed to 1 mm to create a vertical connection. All tests with the DENSE boards were conducted with only two ZA1 connectors. At the end they had been compressed more than 100-times. Nevertheless, they showed no degradation and the springs returned in their initial position.

## 7. Conclusion and Outlook

The existing wafer level redistribution process at the Fraunhofer IZM was successfully extended to enable ultra-high density inter-reticle routing with a pitch of 8 µm at 200 mm wafers. The established post processing technology for 200 mm HICANN wafers is the basis for a large-scale neuromorphic hardware system at the Kirchhoff-Institute for Physics of the Heidelberg University. Based on the BrainScaleS platform each wafer is the core of a neuromorphic module which can be linked together to create large neuronal networks. Up to now a number of 20 BrainScaleS systems are assembled. To enable the upscaling of the network size to a higher number of wafers a different technology platform was evaluated. Instead of discrete assembly and IO connectivity via elastomeric stripe connectors the wafers were embedded into printed circuit boards. This was enabled by adaption of the standard IC embedding technology from Fraunhofer IZM to full wafers.

The embedding of the HICANN wafers in a printed-circuit board was an overall success. It was shown that the embedding process has neither an influence on the circuits nor on the RDLs. There was no degradation or failure during the accelerated environmental stress tests observable. Furthermore, the ZA1 connectors seem to be a reliable connector for later board to board connections.

The next step is to add backside connections to the wafer. They will be used for heat dissipation and later in combination with through silicon vias for the power supply of the wafer. A prototype with backside connections to the silicon is currently in production.

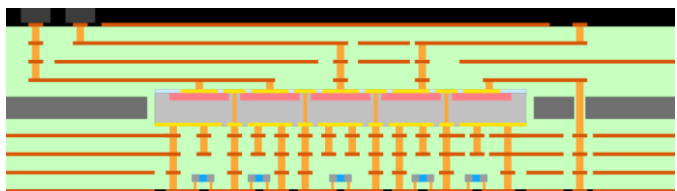

**Figure 15: Possible stack up of the final embedded wafer system with connections on both sides of the silicon wafer, multiple copper layers, TSVs and embedded capacitors.**

After that the number of copper layers will be increased. It should be possible to access half of the HICANNs with an additional copper layer. The long-term goal is to achieve a board like it is shown in Figure 15. It has backside connections, TSVs, multiple copper layers and embedded capacitors. Furthermore, the current 200 mm HICANN wafers are replaced with 300 mm wafers containing the second generation HICANN chips in 65 nm technology.

## Acknowledgments

This work has received funding from the European Union 7th Framework Program ([FP7/2007-2013]) under grant agreements no 604102 (Human Brain Project), 269921 (BrainScaleS) and the Horizon 2020 Framework Program ([H2020/2014-2020]) under grant agreement 720270 (Human Brain Project). The first two authors contributed equally to this work. The authors thank all employees from Kirchhoff-Institute for Physics, Technical University of Berlin and Fraunhofer IZM who have contributed to this work.